\documentstyle[preprint,aps]{revtex}

\begin{document}

\tighten


\title{Significance of the Renormalization Constant of the Color Gauge Field}

\author{M. Chaichian}

\address{ High Energy Physics Division, Department of Physical Sciences\\
University of Helsinki and\\
Helsinki Institute of Physics, P.O. Box 64, FIN-00014,
Helsinki, Finland\\
E-mail: chaichia@pcu.helsinki.fi}

\author{K. Nishijima }

\address{ Nishina Memorial Foundation\\ 2-28-35 Honkomagome,
Bunkyo-ku, Tokyo 113-8941, Japan \\
E-mail: nisijima@phys.s.u-tokyo.ac.jp}

\maketitle

\begin{abstract}
It is shown that a sufficient condition for color confinement is
given by $Z_3^{-1}=0$, where $Z_3$ denotes the renormalization
constant of the color gauge field.
\end{abstract}

\vspace{3ex}


\section{Introduction}

The color degree of freedom has been introduced into the quark
model in order to save the connection between spin and statistics
for quarks and strong interactions are considered to be invariant
under the color $ SU(3)$ group. This new degree of freedom is
hardly recognizable, however, since all the familiar hadrons belong
to the singlet representations of the color $ SU(3)$ group as far
as we are aware. Hadrons belonging to non-singlet representations
of this group, if any, would be realized only at very high
energies. In fact, quarks belonging to the triplet representation
have never been observed to date, thereby suggesting that an
isolated quark is, in principle, not subject to observation. This
is the hypothesis of quark confinement. It indicates one of the
most characteristic features of strong interactions  to favor
neutralization of color by realizing only color singlet states and
suggests a strong resemblance to electromagnetic interactions that
also favor neutralization, though not completely, because of repulsion
between like charges and attraction between unlike charges. This
forms a marked contrast with gravitational interactions which are
always attractive and do not lead to neutralization.

 This analogy led Nambu \cite{nambu1,nambu2} to suspect that strong
 interactions are mediated
 by a gauge field coupled to the color charge. The quanta of the
 color gauge field
 are the color octet gluons. By now it is believed that no
 isolated colored particles are
 observable and we have promoted the hypothesis of quark
 confinement  to that of color confinement.
 In other words, color confinement has
 guided us to the color gauge theory of strong interactions or
 QCD. There is a considerable difference, however, in the degree
 of neutralization between QED and QCD.

  First, let us consider the multipole interactions between
  two electrically neutral systems, which are given by
  the van der Waals potential
\begin{eqnarray}
V_{\rm vdW}(r)\propto r^{-6}.
  \end{eqnarray}
This shows that the electric field exerted by a neutral system can
penetrate into the vacuum without any sharp cut-off.

 The situation is completely different in QCD.
 Let us consider the interaction between
 two color singlet hadrons, for instance,
 nucleon-nucleon scattering. In the language
 of dispersion relations, the potential between them
 is given by the pole contributions in the
 crossed channels. The least massive color singlet particle
 that can be exchanged between
 them is the pion and the resulting nuclear forces
  are represented by the Yukawa potential

\begin{eqnarray}
  V_{\rm Y}(r)\propto \frac{e^{-\mu r}}{r},
  \end{eqnarray}
  where $\mu$ denotes the pion mass. This is a consequence of
  color confinement in that
  isolated quarks and gluons are excluded in the crossed channels.

  By comparing these two  potentials we recognize that the flux of the color
  gauge field exerted  by color singlet nucleons cannot penetrate into the
  confining vacuum as demonstrated by the presence of a sharp cut-off at
 the pion
  Compton wave length as the penetration depth. In this way we realize that
  the Yukawa mechanism of generating nuclear forces leaves a strong
  resemblance to the Meissner effect in the type II superconductor.

  The hypothesis of color confinement also implies
  that the color $SU(3)$ invariance be exact.
  Otherwise, an originally color singlet ground state
  would induce colored states through
  the symmetry-breaking perturbation resulting in the
  leakage of the unconfined color. Thus
  unbroken color gauge symmetry is an important condition
  for confinement and we shall
  take it for granted in what follows.

   So far we have been guided by color confinement to
   reach the concept of color gauge field,
   but we shall also prove the converse that color
   confinement follows from QCD in return. For
   this purpose we first show that a sufficient
   condition for color confinement is given by

\begin{eqnarray}
Z_3^{-1}=0, \label{eq3}
  \end{eqnarray}
where $Z_3$ denotes the renormalization constant of the color
gauge field. Then we prove that this condition is actually
satisfied in a certain class of gauges provided that the color
gauge symmetry is unbroken and asymptotic freedom is respected.


\section{Formulation of QCD}

Let us start from the familiar Lagrangian density for QCD,

\begin{eqnarray}
 L=L_{\rm inv}+L_{\rm gf}+L_{\rm FP},
\label{eq4}
  \end{eqnarray}
where
\begin{eqnarray}
 L_{\rm inv}&=&-\frac{1}{4}F_{\mu\nu}\cdot F_{\mu\nu}
-\overline{\psi}\left(\gamma_\mu D_\mu+m \right)\psi, \\
 L_{\rm gf} &=& A_\mu \cdot \partial_\mu B
+\frac{\alpha}{2}B\cdot B,\\
  L_{\rm FP} &=& i\partial_\mu\overline{c}\cdot D_\mu c.
  \end{eqnarray}
 We have suppressed the color and flavor indices above.
 The first term in  (\ref{eq4})
 is the gauge-invariant piece, the second one
 the gauge-fixing term and the last one the
 Fadeev-Popov ghost term expressed in the
 conventional notation. The Faddeev-Popov ghost
 fields $c$ and $\overline{c}$ are anticommuting quantities.

  The BRS transformations are introduced for the quark
  and gauge fields by replacing the
  infinitesimal gauge function by either $c$ or $\overline{c}$
  in their gauge transformation,
  and they are denoted by $\delta$ or $\overline{\delta}$, correspondingly.
   The BRS
  transformations for other auxiliary fields $B$,  $c$ and $\overline{c}$
  are defined so as to leave the total Lagrangian invariant.
  The conserved BRS charges
  denoted by $Q_{\rm B}$ and $\overline{Q}_{\rm B}$ are
  related to the BRS transformations
  of an operator $F$ through

\begin{eqnarray}
  \epsilon\, \delta F=i[\epsilon Q_{\rm B}, F], ~~~
  \epsilon\, \overline{\delta}
   F=i[\epsilon \overline{Q}_{\rm B}, F],
  \end{eqnarray}
  where $\epsilon$ is a Grassmann variable which anticummutes
  with $c$ and $\overline{c}$. We
  shall skip the BRS transformations of individual fields
  since they are well-known.

  In terms of the BRS transformations the equation for
    the gauge field is given by
\begin{eqnarray}
\partial_\mu F_{\mu\nu}+gJ_\nu=i\delta\overline{\delta}A_\nu,
\label{eq9}
  \end{eqnarray}
where $J_\nu$ denotes the color current density and $g$ the
gauge coupling constant.

The propagation function of the color gauge field is given by the
vacuum expectation value of the time-ordered product of two gauge
fields,

\begin{eqnarray}
\langle A_\mu^a (x),\,A_\nu^b (y)\rangle
=\delta_{ab}\frac{-i}{(2\pi)^4}\int d^4k \,e^{ik(x-y)}D_{\mu\nu}(k),
  \end{eqnarray}
  where $a$ and $b$ are color indices, and
\begin{eqnarray}
D_{\mu\nu} (k)=\left(\delta_{\mu\nu}-\frac{k_\mu
k_\nu}{k^2-i\epsilon}\right)D(k^2) +\alpha\frac{k_\mu
k_\nu}{(k^2-i\epsilon)^2}.
  \end{eqnarray}
  The function $D(k^2)$ obeys the Lehmann representation \cite{leh},
\begin{eqnarray}
D(k^2)=\int dm^2 \frac{\rho (m^2)}{k^2+m^2-i\epsilon}.
  \end{eqnarray}
The renormalization constant of the gauge field $Z_3$ is given by
Lehmann's theorem \cite{leh} by

\begin{eqnarray}
Z_3^{-1}=\int dm^2 \rho (m^2).
\end{eqnarray}


\section{Indefinite Metric and Subsidiary Conditions}

 When a gauge field is quantized in a covariant gauge,
 introduction of indefinite metric
 is inevitable since it is inherited from the Minkowski
 metric through the gauge field
 that transforms as a four-vector. In order to eliminate
 indefinite metric from physical
 interpretations of the gauge theory, subsidiary conditions
 are introduced in order to select
 observable physical states.

  In QED the subsidiary condition called the Lorentz condition is given by

\begin{eqnarray}
  B^{(+)} (x)|\alpha\rangle =0,
  \label{eq14}
  \end{eqnarray}
where $|\alpha\rangle$ denotes a physical state and the
superscript $(+)$ refers to the positive frequency part. In QED
described by the Lagrangian (\ref{eq4}) the Faddeev-Popov ghost
fields $c$ and $\overline{c}$ turn out to be free fields and we
may also require their absence to define a physical state $|\alpha\rangle$
by

\begin{eqnarray}
c^{(+)} (x) |\alpha\rangle =\overline{c}^{(+)} (x) |\alpha\rangle
=0.
 \label{eq15}
  \end{eqnarray}
Thus the subsidiary conditions in QED are given by (\ref{eq14})
and (\ref{eq15}). It is not difficult, however, to verify that the
above conditions are equivalent to those given by (\ref{eq15}) and
(\ref{eq16}):

\begin{eqnarray}
Q_B|\alpha\rangle =0. \label{eq16}
  \end{eqnarray}
We are aware of the fact that (\ref{eq16}) is the only possible
form of the subsidiary condition that can be employed in
non-Abelian gauge theories since the auxiliary fields $B$, $c$ and
$\overline{c}$ are no longer free fields in this case.

\begin{flushleft}
{\bf 4.~Interpretation of Confinement}
\end{flushleft}

In discussing confinement there are two important questions to be
answered. First, we have to answer the question of what
confinement means. Once this question is settled we have to prove
that it is a consequence of QCD in the second stage. In this
section we shall answer the first question.

In order to give an interpretation of confinement, we shall look
for a known example of confinement within the framework of known
theories. We easily find a simple example in QED. Namely,
longitudinal and scalar photons are never subject to observation
and they provide simple examples of confined particles. They are
confined since they fail to satisfy the Lorentz condition or they
belong to zero-norm states. In other words these unphysical
photons are confined by metric cancellation.

By generalizing this argument to QCD we may assume that quarks and
gluons or generally colored particles are confined just because
they fail to satisfy the subsidiary condition (\ref{eq16}),

\begin{eqnarray}
Q_B|\mbox{quark}\rangle \neq 0, ~~~~Q_B|\mbox{gluon}\rangle \neq 0.
\label{eq17}
  \end{eqnarray}
From the definition of the physical states (\ref{eq16}) in QCD we
expect
\begin{eqnarray}
\langle\beta|\delta\overline{\delta}A_\nu (x)|\alpha\rangle =0,
  \end{eqnarray}
when both $ |\alpha\rangle$ and $|\beta \rangle$ are physical. So
we shall show
\begin{eqnarray}
&& \langle\mbox{quark}|\delta\overline{\delta}A_\nu (x)
|\mbox{quark}\rangle {\neq} 0,\nonumber\\
&& \langle\mbox{gluon}|\delta\overline{\delta}A_\nu
(x)|\mbox{gluon}\rangle {\neq} 0,
  \end{eqnarray}
  \label{eq19s}
for the purpose of proving (\ref{eq17}).

By making use of the field equation (\ref{eq9}) and the resulting
Ward-Takahashi identities, we can verify that Eqs.\,(19)
follows from
\begin{eqnarray}
\partial_\mu \langle \delta\overline{\delta}A_\mu^a (x),\,A_\nu^b (y)\rangle = 0.
\label{eq20}
  \end{eqnarray}
This is a sufficient condition for color confinement, but it is
violated when the color symmetry is spontaneously broken
as suggested
in the introduction by an intuitive argument. Also, in
QED Eq.\,(\ref{eq20}) is not satisfied but it is replaced by

\begin{eqnarray}
\partial_\mu \langle \delta\overline{\delta}A_\mu^a (x),\,A_\nu^b (y)\rangle
= \partial_\nu
\delta^4 (x-y), \label{eq21}
  \end{eqnarray}
so that there is no charge confinement.

Furthermore, it can be shown on the basis of renormalization group
and an analysis of the Goto-Imamura-Schwinger term that
Eq.\,(\ref{eq3}) is a sufficient condition for Eq.\,(\ref{eq20}).
Thus the problem of color confinement reduces to that of
evaluating the renormalization constant of the color gauge field
\cite{chai}.


\section{Evaluation of $Z_3$}

 It is worth emphasizing that $Z_3$ can be
  evaluated exactly in QCD by means of renormalization
 group. In the renormalization group approach we study the dependence
 of the parameters characterizing the theory,
 such as the gauge coupling constant $g$ and the
 gauge parameter $\alpha$, on the scale change of
 the renormalization point $\mu$. These
 parameters are
 called the running coupling constant and the running gauge
 parameter as functions of $\mu$
 and are denoted by $\overline{g}$ and $\overline{\alpha}$.
 The asymptotic values of these
 running parameters in the limit of $\mu\to\infty$
 are denoted by $g_\infty$ and
 $\alpha_\infty$, respectively. It should
 be stressed that these asymptotic values can be
 identified with their unrenormalized ones.

  In QED, Gell-Mann and Low \cite{gell} reached the conclusion that
  the unrenormalized
  coupling constant $e^2_\infty/4\pi$ may behave in either of two ways:
  \begin{itemize}
  \item[(a)] It may really be infinite as perturbation indicates;
  \item[(b)] It may be a finite number independent of $e^2/4\pi$.
  \end{itemize}

   In ~gauge ~theories ~such as QED and QCD,
   we ~generally ~have the ~relation \cite{nish}

\begin{eqnarray}
Z_3^{-1}=\frac{\alpha}{\alpha_\infty}.
  \end{eqnarray}
Now asymptotic freedom in QCD is represented by
\begin{eqnarray}
g_\infty=0,
  \end{eqnarray}
  and in this case the asymptotic limit of the
  gauge parameter can assume one of the
  following three alternative values:
  \begin{eqnarray}
  \alpha_\infty=-\infty,\, 0, \, \alpha_0,
  \label{eq24}
  \end{eqnarray}
  where
  \begin{eqnarray}
  \alpha_0=\frac{1}{3}\left(13-\frac{4}{3} N_f\right)
  \end{eqnarray}
  for the system consisting of gluons and $N_f$ flavors of quarks.

   In QED the renormalized coupling constant $e^2/4\pi$
   is fixed  and only one of the two
alternative cases  (a) and (b) is realized. In QCD, however, which
one of the three possible values in (\ref{eq24}) is realized
depends on the initial choice of the two parameters $g$ and
$\alpha$. Although $g$ is fixed as $e$ is, the choice of $\alpha$ is
quite arbitrary. We can always find a domain of $\alpha$ for a
given value of $g$ in which $\alpha_\infty=-\infty$ is realized
and consequently the sufficient condition for confinement
(\ref{eq3}) is satisfied. Indeed, for small values of $g^2$ this
domain is given by \cite{chai,nish}

\begin{eqnarray}
\alpha < \mbox{Min} (\alpha_0,0).
\end{eqnarray}


\end{document}